\documentclass[lettersize,journal]{IEEEtran}
\usepackage{amsmath,amsfonts}
\usepackage{algorithmic}
\usepackage{algorithm}
\usepackage{array}
\usepackage[caption=false,font=normalsize,labelfont=sf,textfont=sf]{subfig}
\usepackage{textcomp}
\usepackage{stfloats}
\usepackage{dirtytalk}
\usepackage{url}
\usepackage{verbatim}
\usepackage{graphicx}
\usepackage{cite}
\begin{document}

\title{Miniature Wireless Power Transfer System \\ for Charging Vertically Oriented Receivers}


\author{
    \IEEEauthorblockN{Andrey Korzin\IEEEauthorrefmark{1}, Esmaeel Zanganeh\IEEEauthorrefmark{1}, Polina Kapitanova\IEEEauthorrefmark{1}}\\
    \IEEEauthorblockA{\IEEEauthorrefmark{1} School of Physics and Engineering, ITMO University, Saint Petersburg, 197101, Russia}
}

\markboth{Journal of \LaTeX\ Class Files,Vol., No., date}%
{Shell \MakeLowercase{\textit{et al.}}: A Sample Article Using IEEEtran.cls for IEEE Journals}

\maketitle

\begin{abstract}
Development of compact wireless power transfer (WPT) systems for charging miniature randomly oriented electronic devices is quite a challenge. Traditionally, WPT systems based on resonant magnetic coupling utilize \say{face-to-face} aligned transmitter and receiver coils providing sufficient efficiency at relatively large distances. However, with the presence of angular receiver misalignment in a such system, the mutual coupling decreases resulting in a low power transfer efficiency. Here we develop a compact WPT system for wireless charging of miniature receivers vertically oriented with respect to the transmitter. As a transmitter, we employ a butterfly coil that provides a strong tangential component of the magnetic field. Thus, a vertically oriented receiver located in the magnetic field can be charged wirelessly. We perform numerical and experimental studies of the WPT system's power transfer efficiency as a function of the distance between the transmitter and the receiver. The misalignment and rotation dependencies of power transfer efficiency are also experimentally studied. We demonstrate the power transfer efficiency of 60~$\%$ within transfer distance of 4~mm for a vertically oriented receiver with an overall dimension of 20~mm~$\times$14~mm at the frequency of 6.78~MHz.

\end{abstract}


\section{Introduction}
\begin{figure}[b]
\begin{center}
\includegraphics[width=6cm]{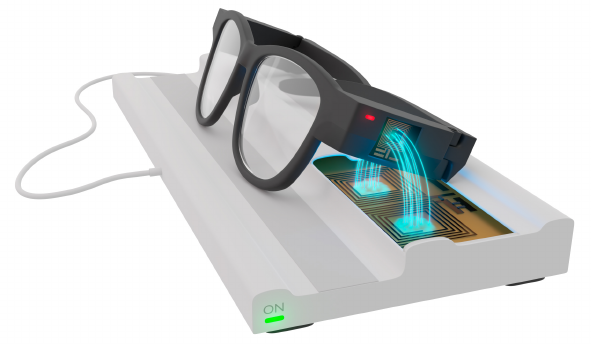}
\caption{\label{fig:1} Conceptual view of the miniature WPT system charging a vertically oriented receiver of a smart glasses.}
\end{center}
\end{figure}

Smart glasses are wearable technology devices that combine the graphic abilities of mobile appliances, artificial intelligence, and augmented reality. With smart glasses, one can make phone calls, get notifications, have 3D map navigation, graphic translation, virtual reality gaming, and more. It is clear that all these utilities 
require some integrated electronics built into these devices, which require a power source, which most often are available in form of batteries. Traditionally, the smart glasses' batteries can be charged via cords or inside of a special case\cite{xu2021}. However, customers always demand more freedom in the use of electronic devices, asking to eliminate cords or cases. From our point of view, a comfortable way of charging these devices would be the positioning of the smart glasses over a charging pad (see Fig.~\ref{fig:1}) where a horizontally oriented transmitter (Tx) coil is incorporated. The smart glasses have to be equipped with a receiver (Rx) coil. Due to the compactness of the device~\cite{matsuhashi2020} itself, there is limited space to locate the Rx coil. Probably the most appropriate way is to incorporate it inside the main part of the glasses' temple, which is not bigger than 85 mm by 20 mm \cite{matsuhashi2020}. Thus, we come to the possible practical scenario when a user places the smart glasses over the charging pad in a spatial position. With the higher probability, we will get the situation when one needs to charge a miniature vertically oriented Rx over the horizontally positioned Tx. However, the angular misalignment is still possible (as shown in Fig.~\ref{fig:1}).

Wireless power transfer (WPT) has been rapidly becoming a significant technology for many applications, such as electric vehicle charging, implanted medical devices, robotic systems, and wearable electronics~\cite{niu2019state, zhang2018wireless, 9316691, Song2021, Zanganeh2023Extreme}. WPT technology based on magnetic resonant coupling has attracted great attention due to its promise of safe and highly efficient mid-range charging \cite{kurs2007wireless, dionigi2022magnetic,song2020multi,zanganeh2021nonradiating,kuzmin2023experimental}. However, the power transfer efficiency (PTE) of magnetic resonance WPT systems with "face-to-face" locations of the Tx and Rx suffers from the change of the Rx position, such as angle and axis misalignment~\cite{7497600,sasatani2021room,kim2016free}.
It can be explained by the decrease in the coupling coefficient between the Tx and Rx while the second one changes its spatial orientation. Moreover, the degradation of the PTE can also be observed for the magnetic resonance WPT systems with the Tx and Rx coils of different overall dimensions. Therefore, a magnetic field shaping technology can be used to control the magnetic flux at a particular location and enhance coupling coefficient, thereby improving positional freedom and PTE simultaneously~\cite{Song2017, zanganeh2022axial}. For that purpose, one can utilize a butterfly Tx coil design~\cite{ha2019} as well as the changing of the current distribution in the planar Tx coils array~\cite{7283671, 2020kang}.

\begin{figure}[t]
\begin{center}
\includegraphics[width=8.3cm]{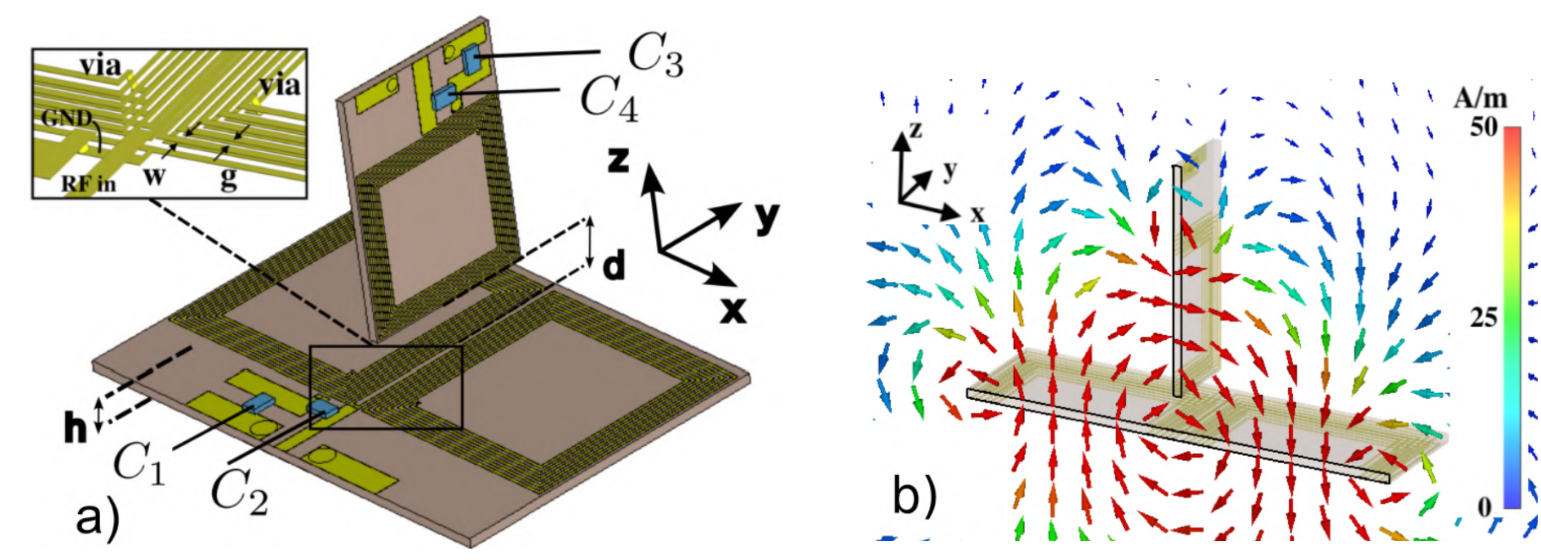}
\caption{\label{fig:2} (a) Schematic view of the miniature WPT system for charging vertically oriented receivers. 
(b) Simulated magnetic field distribution of the WPT system at the frequency of 6.78~MHz numerically obtained for 0.5~W of input power.}
\end{center}
\end{figure}

\begin{figure}[t]
\begin{center}
\includegraphics[width=8.3cm]{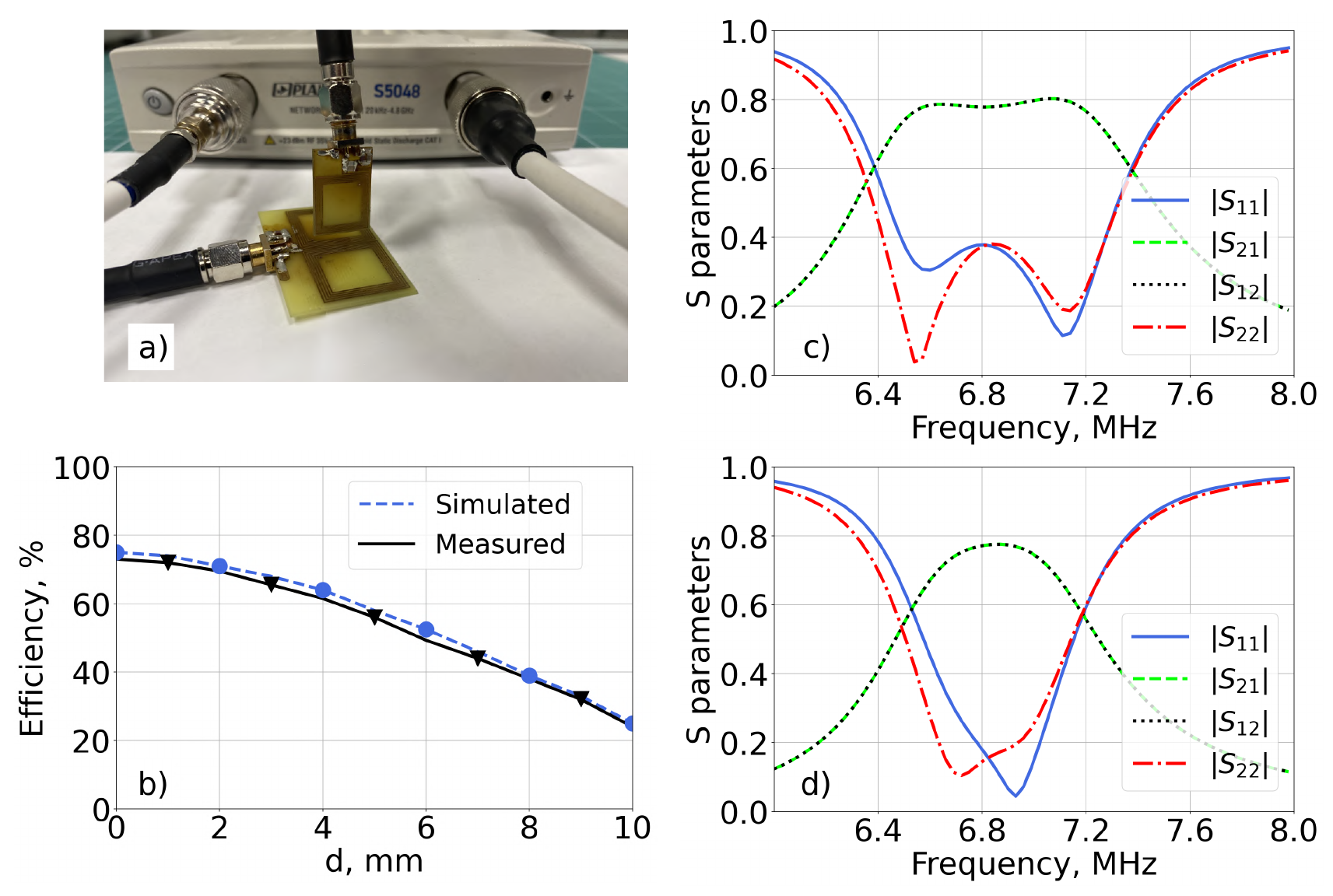}
\caption{\label{fig:3} (a) Photograph of the WPT system prototype connected to a VNA for measurements of the S-parameters. (b) Simulated and measured PTE as a function of the distance $d$. Measured s-parameters for: (c) d = 2 mm and (d) d = 4 mm.}
\end{center}
\end{figure}

\begin{figure*}
\begin{center}
\includegraphics[width=14cm]{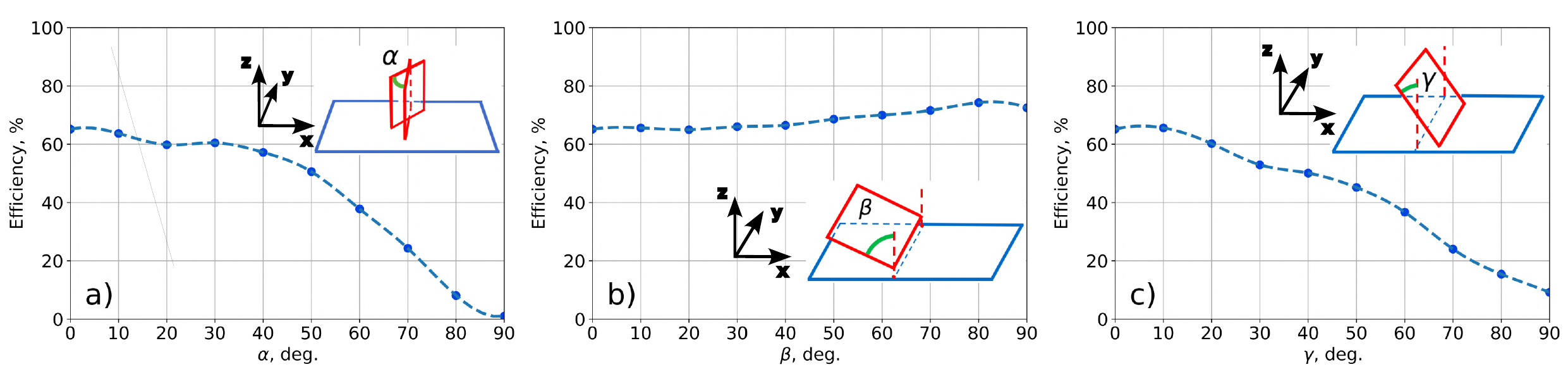}
\caption{\label{fig:4} PTE extracted from the measured S-parameters for different rotation angles of the Rx with respect to Tx: (a) $\alpha$ - rotational angle along the z-axis, (b) $\beta$ - rotational angle along the y-axis with respect to the Rx's lower edge, and (c) $\gamma$ - rotational angle along the y-axis when the Rx center is aligned with the Tx center.}
\end{center}
\end{figure*}

In this paper, we develop a miniature WPT system for charging vertically oriented receivers. We introduce a butterfly-shaped Tx coil wich forms a strong tangential component of the magnetic field, which is used to transfer the power to a vertically oriented Rx coil. We perform numerical simulations of the WPT system in CST Microwave Studio 2022 to study the S-parameters, PTE, and near magnetic field distribution. To prove the numerical simulations, experimental studies of the miniature WPT system are provided in the MHz frequency range. The transfer distances, misalignment, and rotation dependencies of the WPT system PTE are experimentally studied.

\section{WPT System Design}
The miniature WPT system consists of a horizontally located Tx and a vertically oriented Rx separated by a distance $d$, as shown in Fig.~\ref{fig:2}(a). Both Tx and Rx coils are made of 5-turn rectangular copper wires with the thickness of 18~$\mu$m implemented on both sides of FR-4 substrates ($\varepsilon_r$ = 4.5 and $tan(\delta)$ = 0.02) with the height of $h$=0.8~mm. The Tx coil is implemented as a \emph{butterfly coil} with the copper wires winded in opposite directions. The wire width is $w$ = 0.3~mm, and the gap between the wires is $g$ = 0.3~mm. The top and bottom layers of each coil are connected with small copper vias with a radius of $r$= 1.5 ~mm. The overall size of the Tx butterfly coil is 40~mm$\times$27.3~mm. In the Rx, the wire width is $w$ = 0.2~mm, and the gap between the wires is $g$ = 0.2~mm. The overall size of the Rx coil is 20~mm$\times$14~mm. The Tx and Rx coil dimensions are selected to be compatible with smart glasses for possible implementation of the WPT system in the future.

The numerical simulations of the WPT system characteristics are performed in CST Microwave Studio 2022 with the help of the frequency domain solver. The Tx coil is excited through the 50 ohm port, the Rx coil is loaded by 50 ohm port. Due to the opposite wire wingdings of the Tx butterfly coil, the currents flowing in the left and right sides generate a loop of magnetic field penetrating the Tx surface, as shown in Fig.~\ref{fig:2}(b). Over the Tx surface, the tangential component of the magnetic field is dominant. The location of the vertically oriented Rx coil in the maximum of the magnetic field enables a strong coupling and opens a way for power transmission.
The numerically obtained S-parameters for the Tx and Rx coils separation distance of $d$ = 3~mm revealed a poor matching of the WPT system. Thus, we employ matching networks at the input and output of the system. They are based on series-parallel capacitances and indicated in Fig.~\ref{fig:2}(a) as the capacitors $C_1$, $C_2$, $C_3$, and $C_4$. To mount the capacitors, copper contact pads are added at the input and output of the WPT system. During the numerical simulations, we optimize the capacitance values to achieve zero reflection coefficients on both the Tx ($|S_{11}|$) and Rx sides ($|S_{22}|$) at the frequency of 6.78~MHz for the $d$= 3~mm. We reveal the minimums of the reflection coefficients for the following values: $C_1$= 98~pF, $C_2$= 240~pF and $C_3$ = 98~pF, $C_4$ = 170~mm. The PTE of a WPT system can be expressed in term of the S-parameters as follows: 
\begin{equation}
\label{eqn:eq1}
    \eta = \frac{|S_{21}|^2}{1-|S_{11}|^2}\times 100\%
\end{equation}
We perform the numerical simulations of the reflection and transmission coefficients for the different distances $d$ between the Tx and Rx with the matching networks. Using the obtained S-parameters, we calculate the PTE as a function of distance $d$ at the fixed frequency of 6.78~MHz (see Fig.~\ref{fig:3}(b)). The maximum PTE of $77~\%$ is achieved for $d$= 0~mm. For the distance $d$= 3~mm, it reaches $65~\%$. When the distance between the Tx and Rx is increased to $d$=10 mm, the PTE decays to 20$~\%$. It can be explained by the decrease in the coupling coefficient between the Tx and Rx.

\section{Experimental Study of the PTE}
To verify the results predicted numerically, the miniature WPT system prototype is fabricated and demonstrated in Fig.~\ref{fig:3}(a). The Tx and Rx coils are implemented utilizing printed circuit board technology using FR-4 substrates. Surface mounted capacitors with the values obtained during the numerical optimization are mounted on the contact pads. To connect the WPT system input and output to the experimental equipment, SMA connectors are soldered. To measure the S-parameters of the WPT system, a planar Vector Network Analyzer (VNA) S5048 is used. To hold the Rx over the Tx, we use a specially designed holder made of ABS plastic, but it is not shown in Fig.~\ref{fig:3}(a). 

During the experimental investigation, we first measure the S-parameters for separation distance $d$ from Tx to Rx ranging from 0 to 10~mm with a 1~mm step. The measured S-parameters for distances of $d$ = 2~mm and $d$ = 4~mm are shown in Fig.~\ref{fig:3}(c) and (d), respectively. For $d$ = 4~mm, the optimal coupling condition is satisfied, and a single minimum of the reflection coefficient is observed. For $d$ = 2~mm, we observe the frequency splitting phenomenon \cite{sample2011}, which can be explained by a stronger coupling than for the separation distance of $d$ = 4~mm. The measured S-parameters are used to calculate the PTE of the WPT system prototype by Eq.~\ref{eqn:eq1}. The PTE extracted from the measured data is compared with the simulated one in Fig.~\ref{fig:3}(b). A good agreement between the measured data and simulated results is observed.

Next, we mimic the spatial position of the Rx with respect to the Tx. We experimentally study the PTE of the WPT system as a function of the Rx rotational angle with respect to the Tx, as shown in the insets of Fig.~\ref{fig:4}. For the fixed distance $d = 3.5~mm$, we rotate the Rx along the z-axis with a rotational angle $\alpha$ (see the inset of the Fig.~\ref{fig:4}(a)) and measure the S-parameters for angle from $0$ to $90 ^{\circ}$ with a $10 ^{\circ}$ step. The PTE extracted from the measured S-parameters as a function of rotational angle $\alpha$ at the frequency of 6.78~MHz is depicted in Fig.~\ref{fig:4}(a). It is as high as $65 \%$ at $\alpha = 0^{\circ}$ and slightly decreases with increasing of the angle $\alpha$ up to 40$^{\circ}$. With further increase of the angle $\alpha$, it decays faster. The PTE reaches zero for $\alpha = 90^{\circ}$ since the magnetic fields become parallel to the Rx surface and cannot induce current in the Rx coil.
For the fixed distance $d$ = 3.5~mm, we also rotate the Rx along the y-axis with respect to the Rx's lower edge (see inset of Fig.~\ref{fig:4}(b)) with a rotational angle $\beta$ ranging from $0$ to $90 ^{\circ}$ with a $10 ^{\circ}$ step. The experimentally obtained PTE as a function of the $\beta$ is shown in Fig.~\ref{fig:4}(b). As one can see, the PTE is greater than 65~$\%$ regardless of the Rx rotational angle $\beta$. In this rotational scenario, the magnetic field always penetrates the Rx whose lower side is positioned in the center of the Tx and we do not observe the degradation of the PTE.
We also study the situation when the Rx center is positioned along the Tx center and the Rx is rotated along y-axis (see the inset of Fig.~\ref{fig:4}(c)) with the  angle $\gamma$. The obtained PTE is illustrated in Fig.~\ref{fig:4}(c). One can see that the PTE decreases from 64~$\%$ to 10~$\%$ by increasing the $\gamma$ from $0^{\circ}$ to $90^{\circ}$.

\begin{figure}[t]
\begin{center}
\includegraphics[width=8.3cm]{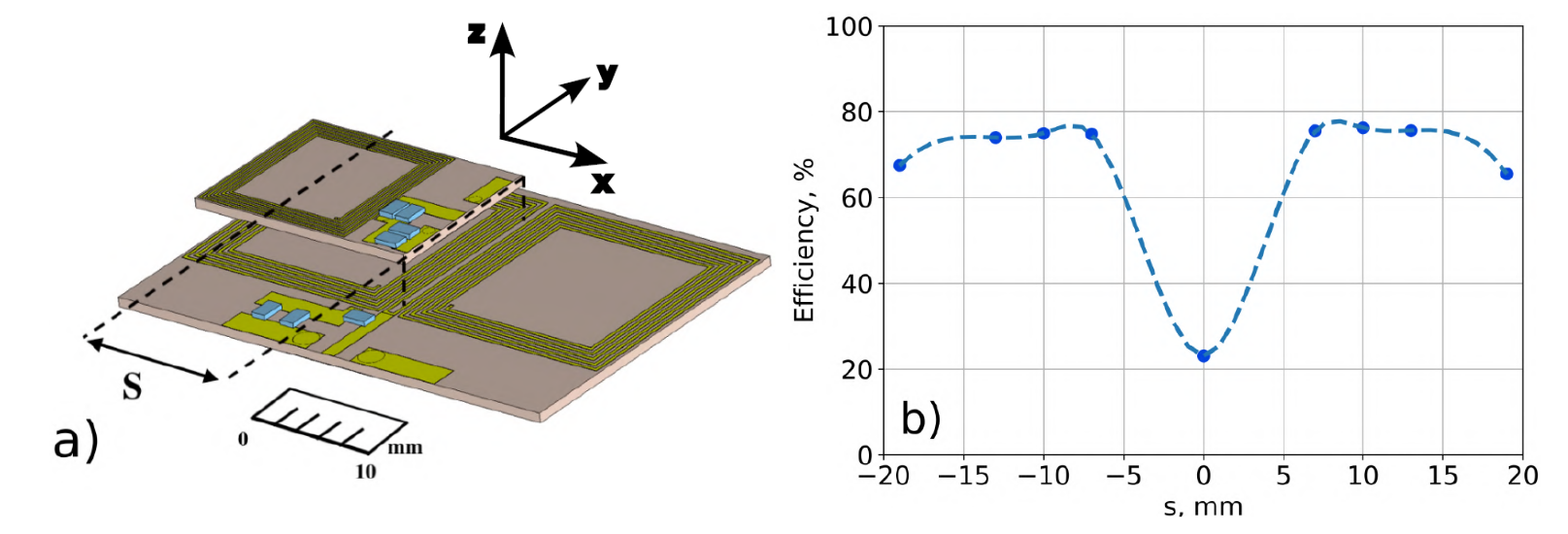}
\caption{\label{fig:5} (a) Schematic view of the WPT system with the Rx parallel to the Tx and a misalignment parameter $S$. (b) Measured PTE as a function of the misalignment parameter $S$.}
\end{center}
\end{figure}

Smart glasses can have many design variations.
Some smart glasses allow you to fully fold the temples and in this case, the Rx coil may be in a horizontal position relative to the Tx butterfly coil. However, the misalignment of the Rx coil regarding the Tx coil may remain. That is why we  also experimentally investigate the PTE dependency for the misalignment of the Rx with respect to Tx, as shown in Fig.~\ref{fig:5}(a). We place the Rx at the separation distance of $d$=3.5~mm parallel to the Tx and move it along the x-axis with a misalignment parameter $S$. Then, we measure the PTE as a function of the misalignment parameter $S$ that varies in the interval $-20 ~ mm \le S \le 20 ~mm $. The PTE extracted from the measured S-parameters at the frequency of 6.78~MHz is higher than 67~$\%$ over the $-20 ~ mm \le S \le -7 ~mm $ and $7~ mm \le S \le 20 ~ mm$ intervals of $S$ since the Rx strongly couples to the butterfly Tx coil. For $S = 0~mm$, the PTE drops to 20~$\%$ due to the compensation of the currents in the Rx coil induced by opposite currents of the Tx coil. The PTE dependence has a slight asymmetry due to the impact of the matching networks of the Rx coil.

\section{Conclusion}
We experimentally studied the PTE of the miniature WPT system composed of the horizontally located Tx and vertically oriented Rx separated by a distance. The transmitter was implemented as a butterfly coil that generates a strong tangential component of the magnetic field, which was used to provide power transfer to the Rx. The overall dimensions of the Tx $40~mm \times 27.3~mm \times 0.8 ~ mm$ and the Rx $20~mm \times 14~mm \times 0.8 ~ mm$ were selected to be compatible with the existing smart glasses. The PTE extracted from the measured S-parameters of the WPT system is higher than 60~$\%$ over transfer distances up to 4~mm for vertically oriented Rx at the frequency of 6.78~MHz. The PTE for rotation with respect to Rx's lower edge is higher than 60~$\%$ from $0^{\circ}$ to $90^{\circ}$. For rotation along the y-axis with respect to the Rx centre, PTE holds higher than 60~$\%$ up to $20^{\circ}$.
Therefore, the transfer distances, misalignment, and rotation dependencies of power transfer efficiency have been successfully demonstrated. The proposed WPT system can be considered as a suitable platform for charging compact devices like smart glasses with vertically oriented receivers.

This work was supported by the Russian Science Foundation (Project No. 21-79-30038)

\bibliography{Main}
\bibliographystyle{IEEEtran}

\vfill

\end{document}